# Single-shot Ultrafast Imaging via Spatiotemporal Division of Femtosecond Laser Pulses

Sarang Yeola, Donghoon Kuk, and Ki-Yong Kim*

*Institute for Research in Electronics and Applied Physics, University of Maryland, College Park, MD 20742, USA*

## Abstract

We have developed a single-shot imaging technique that can capture ultrafast events occurring on femtosecond to picosecond time scales. The technique is based on an optical pump-probe method, in which multiple time-delayed femtosecond pulses simultaneously probe a pump-excited sample. Here we use two sets of 2-by-2 mirror arrays for spatial/temporal division and routing of multiple probe pulses. This single-shot scheme is successfully applied to capture femtosecond ionization fronts propagating at the speed of light in air, as well as laser-induced ablation of solid targets.

Key words: Ultrafast optics; Ultrafast measurements; Ionization

# 1. INTRODUCTION

Since the advent of photography, photographers have been interested in capturing high-speed events. In 1878, Eadweard Muybridge used an array of 16 cameras with each set to capture individual moments of horse galloping, which were then layered in time to create one of the earliest movies [1]. Harold Edgerton, for instance, used stroboscopic flashes to capture a milk drop coronet on a table [2].

Since then the temporal resolution and frame rates of commercial videography has significantly improved, progressing from $10^3$ frames per second (fps) to up to 1 billion fps. For example, Shimadzu has developed a charge-couple device (CCD) that can capture 1 million fps with up to 100 images at resolution of $312 \times 260$ pixels. The frame rates of CCDs and complementary metal-oxide semiconductors (CMOS), however, are fundamentally limited by the time it takes for electrons to register to the digitizer. Frame rates can be further improved when multiple films or cameras are used with a mechanical shutter such as a rotating prism or mirror. For instance, Specialized Imaging in UK produces cameras with framing rates up to 1 billion fps. These and streak cameras typically achieve temporal resolution larger than one picosecond.

A much shorter temporal resolution can be achieved with a femtosecond laser-based pump-probe technique. In this method, an intense laser pulse (pump) excites a sample, and a second pulse (probe) probes the sample after pump's excitation. By capturing and sequencing time-delayed frames, one can reconstruct an ultrafast movie of time-evolving reaction of the sample. This method has been successfully used to capture a superluminal ionization trail produced by an intense laser pulse propagating in air [3] and explosion dynamics of water droplets/jets by femtosecond optical [4] or X-ray [5] laser pulses. This method, however, works for highly repeatable events. If the event to be captured involves irreversible processes such as material damage, chemical reactions, or structural phase transformations, a new sample must be used for each shot under the same laser and sample condition. This often leads to shot-to-shot variations in the condition and thus the resulting dynamics. In such cases, a single-shot method is highly preferable.

Single-shot pump-probe schemes have been also developed. In particular, single-shot spectral interferometry (SSI) can measure both amplitude and phase of a modulated probe, visualizing a pump-induced transient in the complex index of refraction. With chirped supercontinuum light, it can provide a temporal resolution of <10 fs and a >2 ps field of view [6]. This SSI technique has been successfully applied to capture laser-induced double step ionization of helium [7], laser-heated cluster explosion dynamics [8], laser wakefields [9], optical nonlinearity near the ionization threshold [10], and electronic and inertial nonlinear responses in molecular gases [11-14]. This technique, however, is currently limited to provide only 1-dimensional (1-D) spatial information.

Over the past years, many single-shot imaging methodologies have emerged to achieve full 2-dimensional (2-D) spatial imaging with ultrafast temporal resolution. Those include serial time-encoded amplified microscopy (STEAM) [15], spectrally multiplexed tomography (SMT) [16, 17], and sequentially timed all-optical mapping photography (STAMP) [18], spectral filtering STAMP (SF-STAMP) [19], and compressed ultrafast photography (CUP) [20]. All these imaging schemes have their own advantages and disadvantages, as reviewed and compared recently by Mikami *et al* [21]. In general, these techniques require a complicated setup and/or extensive post-processing to retrieve time-resolved frames. Also, in most cases, the achievable temporal resolution is greater than a picosecond, consequently limiting the frame rate less than 1 Tfps.

In this paper, we present a simple single-shot imaging system that utilizes multiple cameras, like Muybridge's camera array technique, to capture ultrafast laser-driven events. Using femtosecond laser pulses, our method can provide a temporal resolution of ~30 fs (corresponding to >30 Tfps) with individually adjustable frame intervals ranging from 30 fs to hundreds of picoseconds.

## 2. METHODS

Our single-shot imaging method is based on an optical pump-probe technique using multiple time-delayed probes to capture the evolution of a pump-excited target in a snapshot. A schematic of our method is shown in Fig. 1. A pulse from a femtosecond laser is split into two pulses (pump and probe). The probe pulse with an initial beam diameter of $D$ is spatially split into four beamlets by a 2-by-2 square-mirror array. Each mirror is slightly angled inward ($\alpha \ll 1$) from 45° so all beamlets cross at a target located $D/(4\alpha)$ away from the mirror array. Each mirror is also set on its own translation stage so that the temporal delay for each beamlet can be individually adjusted. At a 45° angle for the mirror, a 212 µm offset provides a time delay of 1 ps. Given the geometry of our mirror array, each mirror can be translated up to $\sim D/(2\sqrt{2})$ where the reflected beamlet completely misses the target. This provides a maximum possible delay of $D/(2c)$ for each beamlet. Here $c$ is the speed of light in air. For instance, an initial probe beam diameter of $D$ = 45 mm provides a maximum delay range of 300 ps with four mirrors.

After passing the target plane, the probe beamlets are independently reflected by another 2-by-2 square mirror array and directed toward individual cameras through a single imaging lens. Here each mirror is angled outward, with the tilt angle determined by the spacing between the cameras and imaging magnification. The second mirror array is positioned near the focal plane of the lens to provide the best imaging resolution. Note that without the second mirror array, all four images would spatially overlap on the image plane and could not be separated. Also, contrary to the first mirror array, the second one does not need any translation stage for temporal delay offsets. In practice, our experimental setup achieves one-to-one imaging from the target to the cameras using an achromatic doublet of focal length of $f$ = 100 mm.

For experimental demonstrations, we use a Ti:Sapphire amplified laser system capable of producing 800 nm, 4 mJ, 30 fs pulses at 1 kHz. For single-shot experiments, the repetition rate is reduced to 1 Hz. The probe beam is expanded to ~45 mm with 2.5x magnification and directed to the first mirror array. The synchronized pump beam is routed through a delay line and subsequently focused down by a lens ($f$ = 125 mm) at the crossover point of each probe beamlet. The cameras (Imagingsource, DMK 42BUC03) are triggered to capture the images each second and set to be triggered 1.680 ms before the laser sends the pulse. This is the time it takes for the electric signal to reach the cameras and take snapshots under our experimental condition. Finally, the laser pulse is unblocked for less than a second to allow a single-shot experiment.

## 3. EXPERIMENTAL DEMONSTRATION

Our single-shot scheme is first used to capture a femtosecond laser pulse traveling in air. When focused, an intense laser pulse can ionize air molecules, via tunneling and multiphoton ionization, and create an air plasma as it propagates. As the ionization process occurs almost instantaneously, the traveling pulse creates a moving ionization front, like a condensation trail produced by an aircraft, but here propagating nearly at the speed of light. Once formed, the plasma changes the index of refraction of the medium, which in turn modulates the probe's amplitude and phase.

The index of refraction of collisionless and non-magnetized plasma is given by $n = \sqrt{1 - N_e/N_c}$, where $N_e$ is the plasma density, $N_c = \varepsilon_0 m_e \omega^2 e^{-2}$ is the critical plasma density, $\varepsilon_0$ is the electric permittivity of free space, $m_e$ and $e$ is the electron mass and charge, respectively, and $\omega$ is the angular laser frequency. In this experiment, the plasma density ($<10^{17}$ cm$^{-3}$) is much lower than the critical plasma density ($\sim 10^{21}$ cm$^{-3}$) at wavelength of 800 nm. Thus, the probe undergoes mostly a phase modulation with little energy absorption. To visualize the probe's phase modulation, the plasma is set to be formed slightly away from its object plane so that spatial intensity modulations can occur on the image plane due to interference between the phase modulated and unmodulated waves. This phase contrast imaging results in shadowgraphs as shown in Figure 2.

First to synchronize all probe beamlets in time, each mirror is translated so that the corresponding ionization front appears directly over a reference point (see Fig. 2(a)). Also, each mirror-and-camera set is adjusted to line up all air plasma fronts above the same point. Once synchronized, the time delays between the mirrors (or frame intervals) can be freely adjusted as desired. For example, Figures 2(b) and 2(c) show single-shot sequences of shadowgraphs taken with an incremental delay of 250 fs and 1 ps, respectively. It clearly shows that our single-shot camera can capture an ionization front moving at the speed of light with a <30 fs exposure time. We note that each frame, however, provides a different perspective angle (parallax), which is fundamentally caused by our beam splitting scheme. In principle, this stereoscopic effect can be mitigated by rotating each camera (or captured frame) to offset the tilt angle.

A closer look reveals that a faint additional ionization front appears near the ionization front, mostly observed when the pulse reaches near the tip as shown in all Cameras in (a), Cameras 1 and 2 in (b) and (c). This is attributed to an uncontrolled pump pre-pulse that ionizes and creates a plasma prior to the main pulse. This occurs mostly near the tip where the laser is focused and thus has the highest peak intensity.

As a second demonstration, we capture laser-induced plasma creation and expansion on solid surfaces on the timescale of tens to hundreds of picoseconds. In this experiment, the probe pulse is frequency-doubled (400 nm) while the pump wavelength is fixed at 800 nm. This is because a large amount of pump energy is scattered off from the target surface and collected into the cameras, overwhelming the shadowgraph. A 200-µm-thickness beta-barium oxide (BBO) crystal is used to generate second harmonic pulses at 400 nm. An 800 nm mirror is also placed after the BBO crystal to block the original 800 nm component. Finally, an interference filter at 400 nm with 10 nm full width at half maximum (FWHM) is placed in front of each camera to transmit only the frequency-doubled probe beamlet, while blocking the scattered pump pulse, as well as plasma fluorescence light emitted from the target.

Figure 3(a) shows four successive shadowgraphs of a metal wire irradiated by a focused laser pulse, taken at time sequences of 0 ps, 60 ps, 120 ps, and 180 ps. At about 4 ps before the pump pulse reaches the target surface, marked as $t = 0$, there is already a plasma plume with a hemispherical shock boundary on the surface (see Camera 1). It is formed by a pump pre-pulse that arrives 10 ns prior to the main pulse. This is evident from no air ionization trail observed in Camera 1, but present in Cameras 2, 3, and 4. In addition, there is no significant size change observed in the plasma plume and shock boundary over 180 ps. This indicates that the plasma expands on a much longer time scale such as nanoseconds under our experimental condition, also consistent with a previous report [22]. At $t = 60$ ps, the main pump pulse has already interacted with the target surface, leaving another plasma formed right above the preformed one. This is created by additional ionization and heating by the pump laser pulse interacting with ablated nano- and micro-particles and/or the expanding plasma surface where the plasma density greatly varies, and thus the incoming laser energy is greatly absorbed. Similar to the larger plasma, the smaller one does not appear to evolve significantly over a few hundreds of picoseconds. This double plasma formation is a rare event that does not occur always when the experiment is repeated. Perhaps this is due to inconsistent laser focal conditions with respect to the target surface when the target is translated each time so that a fresh surface can be illuminated by the pump pulse. Nonetheless, this demonstrates the usefulness of our single-shot diagnostic in capturing unusual ultrafast events.

Figure 3(b) shows another example of femtosecond laser-induced plasma generation and ablation. This time a 2-μm-thickness aluminum foil is used as a target. At $t = 0$, an intense laser pulse is approaching the target surface as indicated by the air ionization front shown in Camera 1. Here no pre-pulse effect is observed due to a better pre-pulse control. At $t = 60$ ps, the laser pulse has already created a plasma on the surface. At the same time, the impinging laser pulse has been largely reflected from the plasma surface (plasma mirror), creating an air plasma filament in the backward direction as shown in Cameras 2, 3, and 4. As its most energy is reflected, the laser does not appear to penetrate through the thin target. This is also evident from no ionization trace observed on the backside of the target. However, we find that the target is completed drilled through by the laser in a single-shot. This is mostly caused by post plasma heating and ablation occuring on the time scale of hundreds of picoseconds or more as shown in Cameras 2, 3, and 4.

## 4. DISCUSSION

We discuss the capabilities and limitations of our single-shot scheme, especially on its achievable temporal resolution, frame interval, observation time window, and spatial resolution. Firstly, the temporal resolution of each frame (exposure time) is fundamentally determined by the laser pulse duration (~30 fs in our case). In practice, it is additionally constrained by the time it takes for the probe beamlet to transverse the imaging object. Secondly, each frame interval can be freely adjustable, and a current beam size of 46 mm provides a maximum interval of 75 ps per each beamlet. With all probe beamlets equally separated by 30 fs, the maximum frame rate reaches more than 30 Tfps. With all equal intervals of 75 ps, the maximum observation window stretches up to 300 ps. These frame interval and observation window, in principle, can be increased further with a larger input beam and mirrors.

Perhaps the biggest limitation of our scheme arises from spatial resolution. As shown in a ray diagram in Fig. 4, the beamlet (red) probing an object at an angle of $\alpha$ is focused by a lens of focal length $f$ onto the focal plane. This is the plane where the second mirror is located so that all beamlets can be effectively separated and routed to the cameras. This mirror array, however, acts as an aperture, restricting the angle of the imaging cone (blue) from the object. Consequently, this degrades the spatial resolution of each beamlet.

From the diagram, the maximum aperture size is given by $d = 2f \tan \alpha$. This determines the spatial resolution $\delta$ as

$$\delta = \frac{1.22\lambda}{2 \sin \theta} = 0.61\lambda \sqrt{1 + \frac{1}{\tan^2 \alpha}}, \qquad (1)$$

where $\lambda$ is the probe wavelength, and $\theta = \sin^{-1}(1 + f^2/(d/2)^2)^{-1/2}$ is the maximal half-angle of the cone of light. In general, the spatial resolution improves with a large tilt angle $\alpha$ from Eq. (1). With $\alpha = 2°$ in our experiment, the expected resolution is 14 $\mu$m. This resolution is further affected by lens aberration as the beamlet and cone of light does not pass through the center of the lens. When tested with a resolution target, the resolution is estimated to be 20 $\mu$m (see Fig. 5).

In our spatial division scheme, multiple frames are divided from a finite spatial bandwidth. Accordingly, the spatial resolution of each frame degrades with increasing number of frames, while the temporal resolution remains the same. By contrast, in spectral division (or multiplexing) schemes including STAMP, multiple frames are split from a finite spectral bandwidth. In this case, the temporal resolution for each frame worsens with increasing number of frames, whereas the spatial resolution remains unaffected. This sets up a fundamental limit on the space and time resolution product, analogous to the uncertainty principle.

## 5. CONCLUSION

In conclusion, we have developed a single-shot camera system that can capture ultrafast laser-driven phenomena on the femtosecond and picosecond time scales. It has been successfully used to capture femtosecond ionization fronts and ablations on solid targets. In particular, our camera system could capture double plasma formation and plasma mirror generation on solid targets. These interesting features are not easily observable with conventional multi-shot pump-probe schemes especially when the laser and target conditions vary with large shot-to-shot fluctuations.

The utmost advantage of our scheme lies on its easy implementation and capability of providing freely adjustable frame intervals ranging from tens of femtoseconds to several hundreds of picoseconds. One fundamental limitation, however, is that our camera system provides stereoscopic views with limited spatial resolution. Nonetheless, our method can be improved further. The number of frames per shot is currently limited to 4 but can be increased further with more splitting mirrors. In addition, an optical interferometer can be inserted between the second mirror array and camera sets to yield single-shot, time-resolved, 2D interferograms. This will be useful in visualizing time-varying plasma densities, femtosecond relativistic electron beams, and many more in ultrafast laser-matter interactions.

**Funding**. National Science Foundation (NSF) (1351455).

**Figure Captions**

**Fig. 1.** Schematic of our single-shot imaging technique based on spatial/temporal division and routing. The incoming probe pulse is projected onto a 2-by-2 mirror array, which splits the probe into four beamlets. Each mirror is tilted inward and translated with an offset so that all four beamlets probe the sample with different temporal delays. Another set of 2-by-2 mirror array with each mirror angled outward is used to route the beamlets to individual cameras with a single imaging lens.

**Fig. 2.** Single-shot sequences of a femtosecond laser-induced ionization front propagating in air, taken with incremental time delays of (a) 0 fs, (b) 250 fs, and (c) 1 ps. All probe delays are clocked at $t = 0$ when each imaged ionization front passes directly above the sharp tip.

**Fig. 3.** Single-shot sequences of femtosecond laser-induced plasma creation and ablation on the surface of (a) a metal wire and (b) a 2-μm-thickness aluminum foil, both exposed in atmospheric air. In (a), the laser pulse creates an additional plasma on top of the plume pre-formed by a laser pre-pulse. In (b), the impinging laser pulse ablates the target and simultaneously reflects from the surface, making an ionization trail in air.

**Fig. 4.** Ray diagram representing one of four beamlets (red) that passes through the objective lens, plus the cone of light (blue) produced from a point on the object to the image plane through an aperture introduced by the mirror array on the focal plane.

**Fig. 5.** The 1951 USAF resolution test chart imaged by individual cameras in our single-shot imaging system. The resolution here determined by the chart is ~20 $\mu$m.

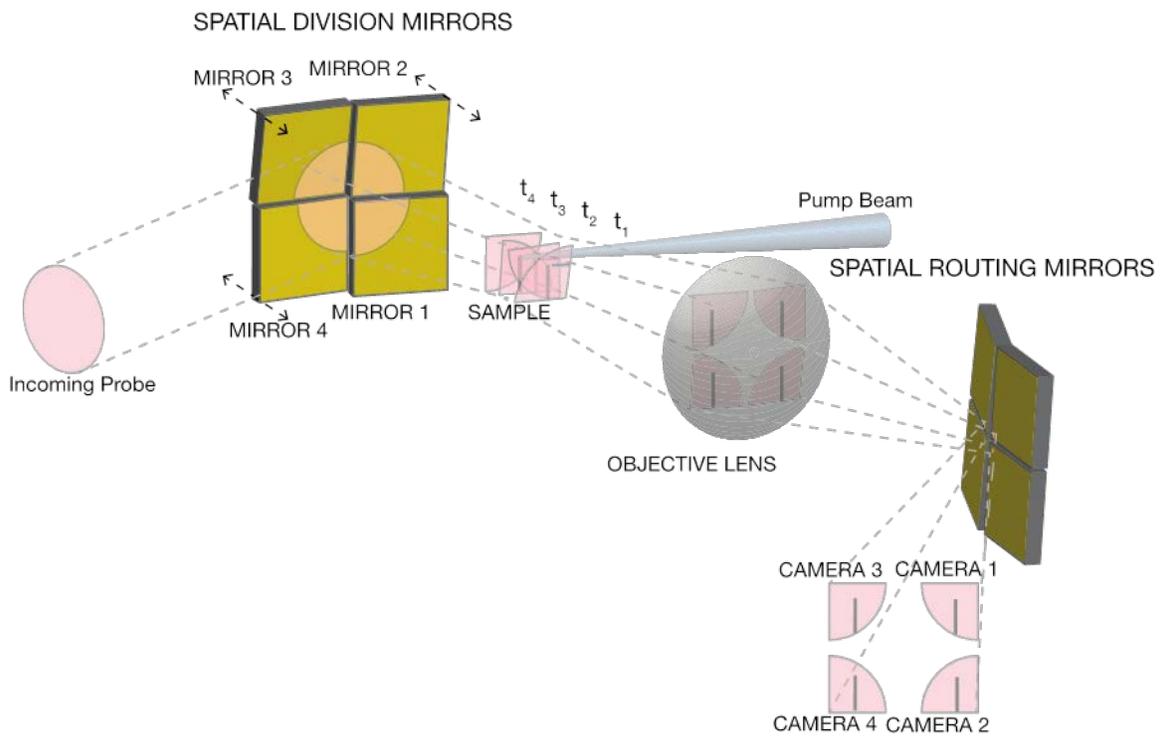

**Figure 1**

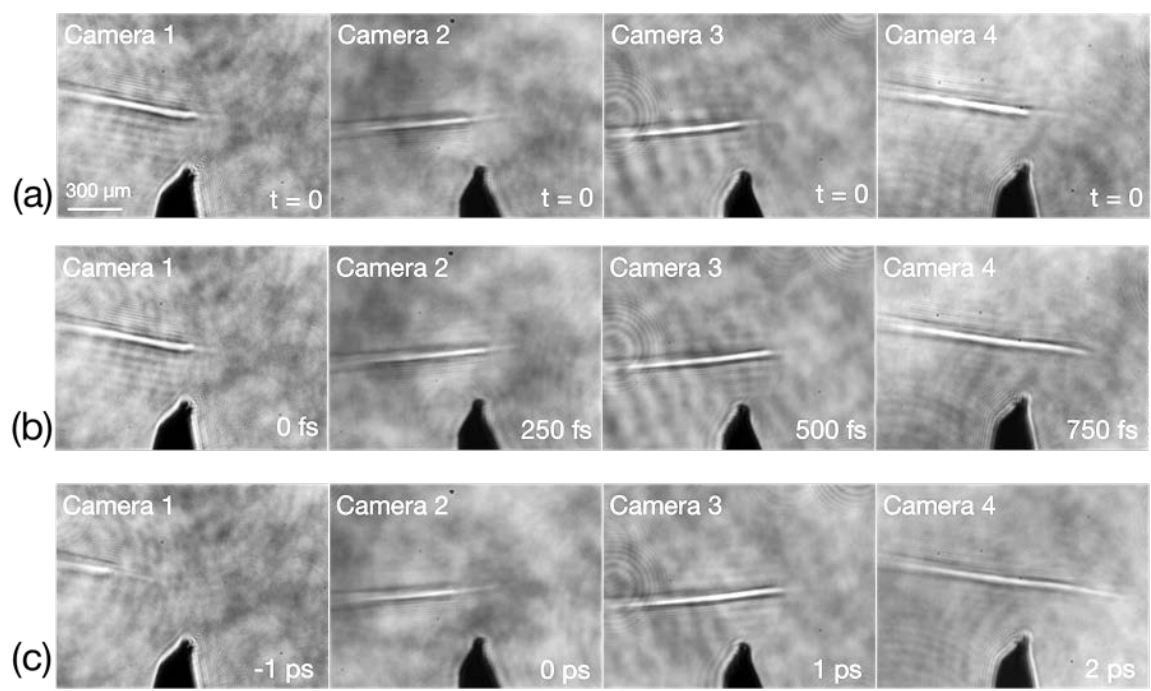

**Figure 2**

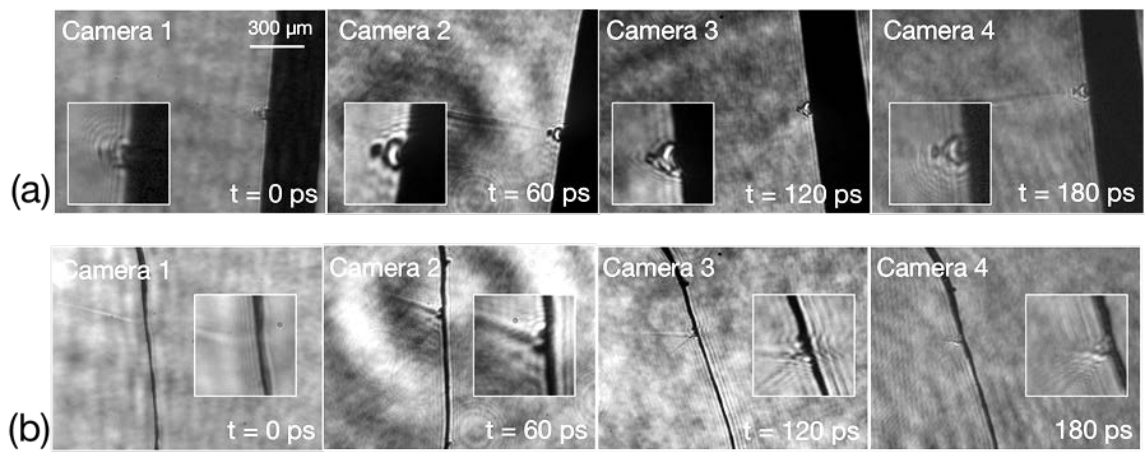

**Figure 3**

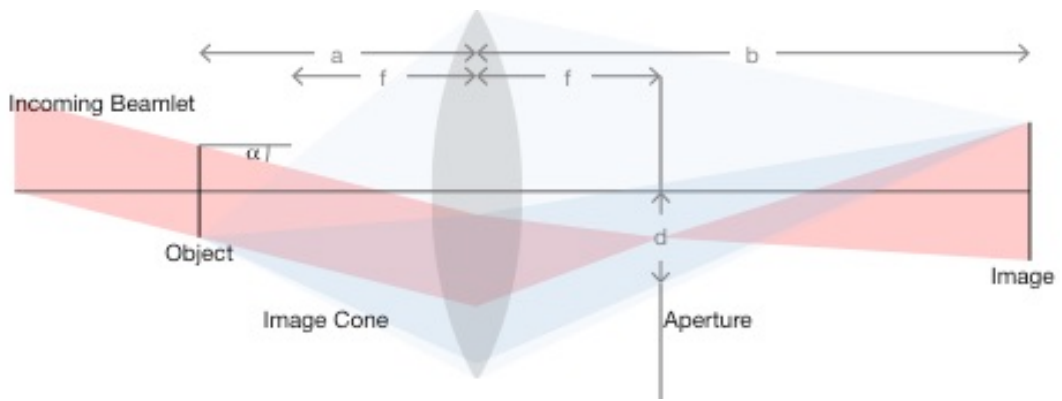

**Figure 4**

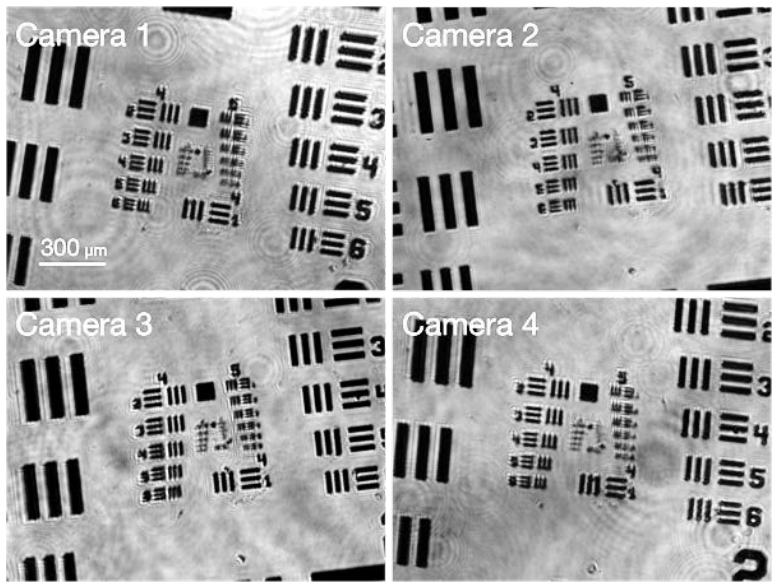

**Figure 5**